\documentclass[pra,twocolumn,amsmath,amssymb,groupaddress,eqsecnum]{revtex4-2}
\usepackage{graphicx,amsmath,relsize,epstopdf,color,mathtools,bm,newtxtext,newtxmath,braket,rotating}
\usepackage[hyphenbreaks]{breakurl}
\usepackage[colorlinks=true,linkcolor=blue,citecolor=blue,urlcolor =blue]{hyperref}
\usepackage[normalem]{ulem}
\usepackage[table,xcdraw]{xcolor}
\usepackage{braket}

\begin{document}
\title{On the ray-wavefront duality}

\author{Antonio D\'{\i}az-Cano} 
\affiliation{Departamento de Algebra, Geometr\'{\i}a y Topolog\'{\i}a,  Universidad Complutense, 28040~Madrid, Spain}
\affiliation{Instituto de Matem\'atica Interdisciplinar, Universidad Complutense, 28040~Madrid, Spain}

\author{Francisco  Gonz\'{a}lez-Gasc\'on}
\affiliation{Departamento de F\'{\i}sica Te\'orica, Universidad Complutense, 28040 Madrid, Spain}

\author{Luis L. S\'{a}nchez-Soto}
\affiliation{Departamento de \'Optica, Universidad Complutense, 
28040 Madrid, Spain}
\affiliation{Max-Planck-Institut f\"{u}r die Physik des Lichts, 91058 Erlangen, Germany}

\begin{abstract}
We investigate the behavior of the solutions of the ray equation in isotropic media. We discuss local and global transversals and wavefronts and give examples of rays without wavefronts.
\end{abstract}

\maketitle

\section{Introduction}

Geometrical optics is a peculiar science. Its fundamental ingredients are rays, which do not exist (except as a mathematical idealization), and wavefronts, which indeed do exist, but are not directly observable~\cite{Stavroudis:2006fk}. Yet it works: even with such an unsophisticated background, it maintains a unique position in modern technology~\cite{Iizuka:2008uq}.

Superficially, geometrical optics might appear as a naive picture of light propagation. However, the seminal work of Luneburg~\cite{Luneburg:1966ly} and Kline and Kay~\cite{Kline:1965qf} laid the solid foundations of this discipline: the wavefronts come associated with the eikonal equation, which is a short-wavelength approximation to Maxwell's equations.  With a small amount of calculation, one can also show that the rays are normal to the wavefronts. It is thus not surprising that the authoritative textbook by Born and Wolf~\cite{Born:1999fk} states that ``only normal congruences are of interest for geometrical optics''. Consequently, non-normal congruences are safely ignored, with a few exceptions~\cite{Luneburg:1966ly,Synge:1937uq} that look at them more as an exotic curiosity than as a \mbox{feasible possibility.}

Put in a slightly different manner, a perfect duality between rays and wavefronts is tacitly assumed, so both can be used interchangeably.  Indeed, given the shape of a wavefront, the direction of any ray crossing the wavefront can be immediately calculated via the eikonal equation. Conversely, one might hope that the shape of a wavefront can be calculated by tracing enough rays.

However, a closer examination reveals that this latter belief is not fully justified.  We revisit here that problem: taking the ray equation as our starting point, we address the simple and unexplored question  of whether given a family of rays one can always find the associated wavefronts. The unforeseen answer we find is that while in a two-dimensional world, the wavefronts always exist (so the ray-wavefront duality is correct), it is not generally the case for three-dimensional rays.


\section{The Ray Equation}

To be as self-contained as possible, we first briefly summarize the essential ingredients that we shall need for our purposes.  In geometrical optics, light propagates along rays, which are taken as oriented curves whose direction coincides everywhere with the direction of the propagation of the energy (i.e., the average Poynting vector). 

Let $\mathbf{x} (t) \in \mathbb{R}^{m}$ (with $m=1, 2,$ or 3 being the space dimensionality) denote the position vector of a point on a ray, considered as a function of an arbitrary parameter $t$, which can be thought of as time.  These curves can be obtained via Fermat's principle~\cite{Caratheodory:1937zr}; that is, as the variational problem $\delta \mathcal{A} = 0$~\cite{Elsgolts:2003uq}, where
\begin{equation}
  \label{eq:opl}
  \mathcal{A} = \int_{t_{1}}^{t_{2}} L(\mathbf{x}, \dot{\mathbf{x}} )  \; dt 
\end{equation}
and the optical Lagrangian is~\cite{Lakshminarayanan:2002wx}
\begin{equation}
L(\mathbf{x}, \dot{\mathbf{x}} ) = n(\mathbf{x}) \, \lVert \dot{\mathbf{x}} \rVert \, .
\end{equation} 

Here, $n (\mathbf{x})$ is the refractive index,  the dot indicates the derivative with respect to $t$, and $\lVert \cdot \rVert =$ stands for the Euclidean norm. Notice that we are assuming that light propagates in an isotropic nondispersive medium.

This is a standard  problem in the calculus of variations and the time-honored Euler--Lagrange equations (which give a sufficient condition of extremality) reduce in this \mbox{case to }
\begin{equation}
\label{eq:rayeq}
n  \,  \ddot{\mathbf{x}} =  \dot{\mathbf{x}}^2 \, {\nabla} n - 
2 \dot{\mathbf{x}} \, {\nabla} n \cdot \dot{\mathbf{x}}  \, ,
 \end{equation}
which is called the ray equation. Quite often, this equation is rewritten in the form
\begin{equation}
\frac{d}{ds} \left ( n \frac{d\mathbf{x}}{ds} \right ) = \nabla n \, ,
\end{equation}
in terms of the Euclidean arc-length parameter $s$, such that $ds = \lVert \dot{\mathbf{x}} \rVert \, dt$. 

The function 
\begin{equation}
I(\mathbf{x}, \dot{\mathbf{x}}) = n^{2}(\mathbf{x}) \, \dot{\mathbf{x}}^{2} 
\end{equation}
is a first integral of~\eqref{eq:rayeq}, as can be checked by observing that
\begin{equation}
\dot{I} (\mathbf{x}, \dot{\mathbf{x}})  = \frac{\partial I}{\partial  \mathbf{x}}  \, \dot{\mathbf{x}} + 
\frac{\partial I}{\partial \dot{\mathbf{x}}}  \, \ddot{\mathbf{x}}  = 0 \,  ,
\end{equation}
where $\partial /\partial \mathbf{x}$ denotes here the gradient with respect to $\mathbf{x}$ and analogously for $\partial / \partial \dot{\mathbf{x}}$. 

Since for light in isotropic media $n (\mathbf{x}) = 1/\lVert \dot{\mathbf{x}} \rVert$ (taking, for definiteness, the speed of light in vacuum as 1), only the level set of $I=1$ is of optical interest.

On the other hand, the Hamilton--Jacobi equation~\cite{Arnold:1989qf} associated to the extremal problem \eqref{eq:opl} is the eikonal equation:
\begin{equation}
\label{eq:eik}
(\nabla \mathcal{S} )^{2} = n^{2} (\mathbf{x}) \, ,  
 \end{equation} 
a term coined by Bruns as early as 1895~\cite{Bruns:1895ti}. This equation can alternatively be obtained as an asymptotic limit (for short wavelengths) of Maxwell's equations~\cite{Born:1999fk}: the real scalar function $\mathcal{S} (\mathbf{x})$ represents the optical path for a locally plane wave. 

{The characteristics associated to the first-order partial differential equation~\eqref{eq:eik} \mbox{are~\cite{Kravtsov:1990kx,Siddiqi:1999ve,Slawinski:2011vn}}}
\begin{equation}
\label{eq:Lieder1}
\dot{\mathbf{x}} = \frac{\nabla \mathcal{S}}{(\nabla \mathcal{S})^{^2}} \, .
 \end{equation} 

This shows that the rays are orthogonal to the level sets of $\mathcal{S} (\mathbf{x})$, which are called wavefronts. The derivative along the streamlines of the vector field $\frac{\nabla \mathcal{S}}{(\nabla \mathcal{S})^{^2}}$ is
\begin{equation}
\label{eq:wavfro}
\dot{\mathcal{S}} = \nabla \mathcal{S} \cdot \dot{\mathbf{x}} = 1 \, , 
\end{equation}
which ensures that the level sets $\mathcal{S} (\mathbf{x}) = C$ are transported by the rays $\mathbf{x}$.

The vector field $\frac{\nabla \mathcal{S}}{(\nabla \mathcal{S})^{^2}}$ is complete (i.e., its flow is defined $\forall t \in \mathbb{R}$) since
\begin{equation}
\left \lVert \frac{\nabla \mathcal{S}}{(\nabla \mathcal{S})^{^2}} \right \rVert = \frac{1}{n} \le 1 \, .
\end{equation}

This vector field defines the symmetry group of the level sets $\mathcal{S} (\mathbf{x}) = C$ and this, in turn, implies that  the level curves of $\mathcal{S} (\mathbf{x})$ are topological straight lines (or planes). This is confirmed by a direct integration of \eqref{eq:wavfro}: $\mathcal{S} (t) =  \mathcal{S}_0 + t$. These conclusions are no longer valid, however, when \eqref{eq:rayeq} is only satisfied locally in a region $\Omega \subset \mathbb{R}^{m}$; for example, when $\mathcal{S}$ vanishes at some points. In that case, the level sets of $\mathcal{S}$ can have many interesting forms.

It is straightforward to check that all the solutions of \eqref{eq:Lieder1} do satisfy the ray equation~\eqref{eq:rayeq}. Therefore, once we know the wavefronts, the rays can be always directly determined: this is the backbone of the ray--wavefront duality. 

If instead of the vector field $\frac{\nabla \mathcal{S}}{(\nabla \mathcal{S})^{^2}}$  we consider a smooth field $\mathbf{X}$, the question arises of whether or not functions $n(\mathbf{x})$ and $\lambda (\mathbf{x})$ can be found such that all the solutions of
\begin{equation}
\label{eq:Lieder2}
\dot{\mathbf{x}} = \lambda (\mathbf{x}) \, \mathbf{X} (\mathbf{x}) 
 \end{equation}  
satisfy the ray equation~\eqref{eq:rayeq}. A sufficient condition can be directly found by introducing $\dot{\mathbf{x}} = \mathbf{X}/(n \lVert \mathbf{X} \rVert )$ into~\eqref{eq:rayeq}, getting
\begin{equation}
\frac{d}{dt} \left ( \frac{\mathbf{X}}{n \, \lVert \mathbf{X} \rVert} \right ) +
\frac{2 \mathbf{X} (\nabla n \cdot  \mathbf{X})}{n^{3} \, \lVert X \rVert^{2}} - \frac{\nabla n}{n^{3}} = 0 \, .
\end{equation}

This constitutes a first-order set of partial differential equations for $n (\mathbf{x})$ that, in general, has no solutions for $n(\mathbf{x})$ when $\mathbf{X}$ is given.

\section{Rays in $\mathbb{R}^{2}$}

In this section, we restrict our attention to the case of two-dimensional vector fields $\mathbf{X}$. Then, we can show that a function $\lambda (\mathbf{x})$ can be always found such that locally we have
\begin{equation}
\label{eq:locfield}
\lambda (\mathbf{x}) \, \mathbf{X} (\mathbf{x}) = \frac{\nabla \mathcal{S}}{(\nabla \mathcal{S})^{2}} , 
\end{equation}
inside a Euclidean ball $B_{0}$ of center $P_{0}$ [with $\mathbf{X} (P_{0}) \neq 0$]. The proof is simple. Let $\mathbf{X} = (X, Y)^{\top}$ (the superscript $\top$ being the transpose) be a smooth  field and $\mathbf{X}_{\perp} = (- Y, X)^{\top}$. Let $\mathcal{S} (x,y)$ be a local first integral of $\mathbf{X}_{\perp}$ in a neighborhood of the point $P_{0} \in \mathbb{R}^{2}$ such that $\mathbf{X}_{\perp} (P_{0}) \neq 0$ and $\nabla \mathcal{S} (x,y)|_{P_{0}} \neq 0$. Then, \eqref{eq:locfield} defines $\lambda$, since by construction $\mathbf{X}$ and $\nabla \mathcal{S}$ are parallel in the ball $B_{0}$. 

Some remarks seem in order here. Assume that the vector field $\lambda \mathbf{X}$ has orbits near $P_{0}$ that are bounded for $t > 0$. This is fulfilled when the orbits are compact (case of cyclic orbits) or tend to a compact set $K$ for $t>0$. Then, the solutions $\lambda \mathbf{X}$ satisfy the ray equation on $B_{0}$, and since they are analytic functions of $t$ for any $t \geq 0$, by prolongation, the ray equation will also be satisfied for any $t$. Therefore, orbits of plane vector fields tending to an equilibrium point~\cite{Guckenheimer:1983up} or a limit cycle can be considered, conveniently parametrized, as solutions of the ray equation.

We recall that Thom's theorem~\cite{Kurdyka:2000ug} implies that all the orbits of a $\mathbb{R}^{2}$ analytic vector field $\mathbf{X}$ of type gradient cannot spiralize around an equilibrium point of $\mathbf{X}$. This means that near a focus of $\mathbf{X}$, \eqref{eq:locfield} does not hold, but it holds near a node. Note that the equilibrium point itself is not a solution of the ray equation, as the constraint $n^{2} ( \mathbf{x}) \,  \dot{\mathbf{x}}^{2} = 1$ is not satisfied when $n(\mathbf{x})$ is a global function of $\mathbf{x}$.

We conclude this section with two \emph{global} results on wavefronts for nonvanishing $\mathbb{R}^{2}$ vector fields. Assume first that a conformal (i.e., angle-preserving) diffeomorphism $D$ can be found such that $D(\mathbf{X}) = \frac{\partial}{\partial x}$. Then $D^{-1} (x=C)$ and $D^{-1} (y=C)$ are first integrals of $\mathbf{X}_{\perp}$ and $\mathbf{X}$, respectively.

Let $\mathbf{X}_{\perp}$ be a nonvanishing divergence-free vector field. Then, the global function $\mathcal{S}$ satisfying $\mathbf{X}_{\perp} = (\mathcal{S}_{x}, - \mathcal{S}_{y})^{\top}$ is computable by quadratures and is a global first integral of $\mathbf{X}_{\perp}$. Therefore, $\mathcal{S} (x,y) = c$ are wavefronts of $\mathbf{X} = \frac{\nabla \mathcal{S}}{(\nabla \mathcal{S})^2}$ since $\dot{\mathcal{S}} = 1$. The divergence-free hypothesis is essential; actually, Wazewski~\cite{Wazewski:1934vp} constructed examples of nonvanishing smooth vector fields $\mathbf{X}_{\perp}$ free from nontrivial global first integrals.

\section{Rays in  $\mathbb{R}^{3}$}
We turn our attention to the three-dimensional vector fields reducible to the type \eqref{eq:Lieder1}. In general, the eikonal defining the wavefronts does not exist even locally.  To guarantee the existence of local orthogonal surface $\mathcal{S} (x,y,z)= C$ to the vector field $\mathbf{X} (x, y, z)$, we require  
\begin{equation}
\label{eq:locint}
X dx + Y dy + Z dz = 0 \,, 
\end{equation}
which can be recast as
\begin{equation}
\label{eq:inte}
dz = U \, dx + V \, dy \, ,
\end{equation}
with $U = - X/Z$ and $V = - Y/Z$. The integrability condition of \eqref{eq:inte} reads
\begin{equation}
  \label{eq:Xcon3}
  \frac{\partial U}{\partial y} + \frac{\partial U}{\partial z} V =
  \frac{\partial V}{\partial x} + \frac{\partial V}{\partial z} U \, .
\end{equation}

When~\eqref{eq:Xcon3} is satisfied, we get the function $\mathcal{S} (x,y,z)$ from \eqref{eq:inte}, such that $\nabla \mathcal{S}$ is parallel to $\mathbf{X}$. Note that equations  \eqref{eq:inte} and~\eqref{eq:Xcon3} are invariant under the replacement $\mathbf{X} \mapsto \lambda (x,y,z ) \mathbf{X}$. 

The vector fields of the type $\mathbf{X} = \nabla \Phi$, where $\Phi$ is a smooth scalar function, satisfy these equations automatically, and the same happens for two-dimensional vector fields, writing $dx = - X/Y dy$ instead of \eqref{eq:inte}.

The vector fields orthogonal to $\mathbf{X}$ lie on the level sets of $\mathcal{S}$. According to \eqref{eq:Xcon3}, these vector fields form an integrable distribution of dimension 2~\cite{Boothby:2003wz}, $\mathcal{S}$ being a first integral \mbox{of it.}

Let us explore the situation with some examples. Let $\mathbf{X}$ be 
\begin{equation}
\label{eq:ex1}
\mathbf{X} = \begin{pmatrix}
y^{2} \\
z \\
- y
\end{pmatrix} \, . 
\end{equation}

One can immediately check that this field satisfies the integrability constraint \eqref{eq:Xcon3}. Equation~\eqref{eq:inte} takes now the form (for $z$ constant)
\begin{equation}
y \, dx + \frac{z}{y} \, dy = 0 \, , 
\end{equation}
whose solution, apart from an additive constant, reads
\begin{equation}
x = \frac{z}{y} \, .
\end{equation}

Therefore, we conclude that the family of surfaces $x - z/y = C$ are locally orthogonal to $\mathbf{X}$. In consequence, the rays are determined by  $\nabla \mathcal{S} = (1, z/y^{2}, - 1/y)^{\top}$, and the resulting refractive index is
\begin{equation}
n (x, y, z) = \sqrt{1 + \frac{1}{y^{2}} + \frac{z^{2}}{y^{4}}} \, ,
\end{equation} 
which is not defined on $y=0$. Nonetheless, the vector field \eqref{eq:ex1} on $y=0$ is orthogonal to the singular plane $y=0$. Note that $\dot{\mathcal{S}} = \nabla \mathcal{S} \cdot \mathbf{X} \neq F (\mathcal{S})$ so \emph{global} wavefronts of $\mathbf{X}$ cannot\mbox{ be obtained.}

 \begin{figure*}
 \includegraphics[height=5cm]{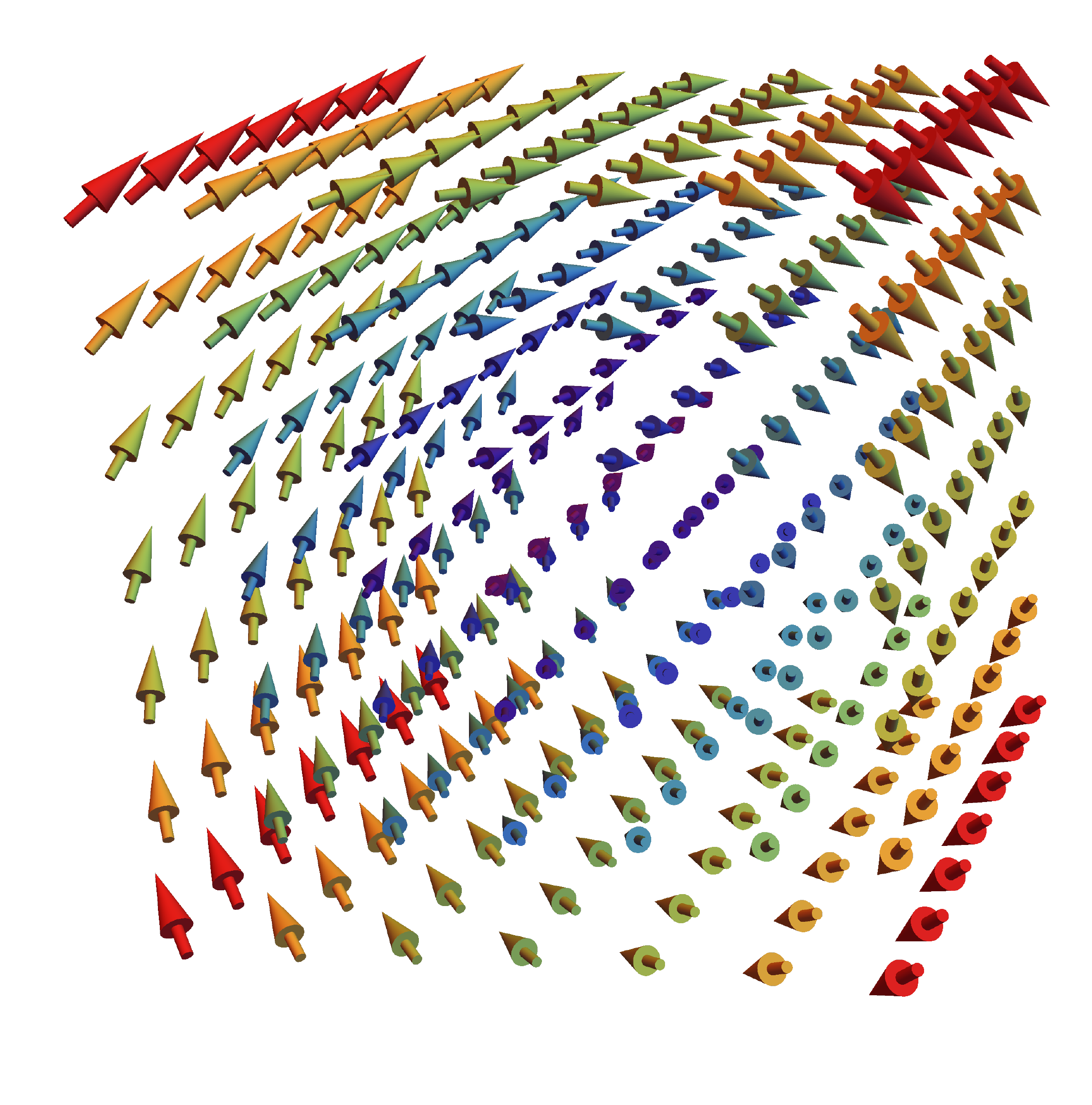}
\qquad \qquad
\includegraphics[height=5cm]{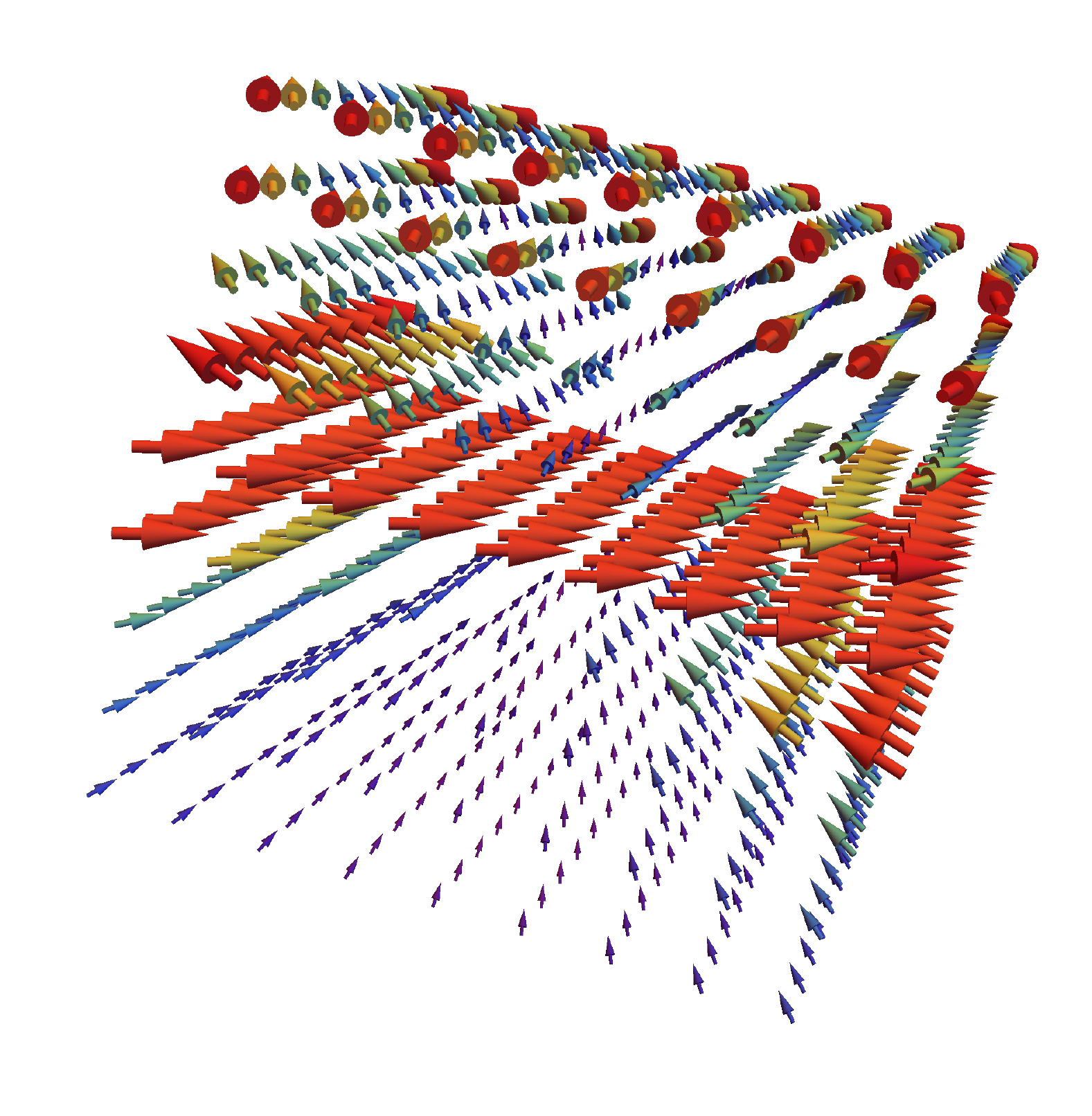}
\caption{\label{schematic} Plots of the vector fields \eqref{eq:ex3} (\textbf{left}) and \eqref{eq:ex4} (\textbf{right}).}
\label{fig:vf}
\end{figure*}

As a second example, let $\mathbf{X}$ be 
\begin{equation}
\label{eq:ex2}
\mathbf{X} = \begin{pmatrix}
- y \\
x \\
- \lambda (x^{2} + y^{2})
\end{pmatrix} \, . 
\end{equation} 

The integrability condition \eqref{eq:Xcon3} holds when $\lambda (u) = C u$, where $C \in \mathbb{R}$. If, for simplicity, we fix $C=1$,~\eqref{eq:locint} becomes
\begin{equation}
- y \, dx + x \, dy + (x^{2} + y^{2}) dz = 0\, ,
\end{equation}
with solution give a local family of transversals to $\mathbf{X}$:
\begin{equation}
\mathcal{S}(x,y,z) = z  - \arctan \left ( \frac{x}{y}\right ) \, .
\end{equation}

The  wavefront condition $\dot{\mathcal{S}} = 1$ is  not satisfied because $\dot{\mathcal{S}} = \nabla \mathcal{S} \cdot \mathbf{X} = 1 + x^{2} + y^{2}$, and accordingly the vector field does not admit wavefronts.

Let now $\mathbf{X}$ be the vector field
\begin{equation}
\label{eq:ex3}
\mathbf{X} = \begin{pmatrix}
- y \\
x \\
1
\end{pmatrix} \, , 
\end{equation} 
whose orbits are the helices
\begin{eqnarray}
  \label{eq:helicex}
  x (t)=  a \,  \cos (t + \varphi) \,  ,
  \quad
  y (t) =  a \, \sin (t + \varphi) \, ,
  \quad 
  z (t) =  t \,  ,
\end{eqnarray}
with $a , \varphi \in \mathbb{R}$, $a \ge 0$. In Fig.~\ref{fig:vf}, we plot the integral curves of this vector field, showing an intriguing chiral behavior. One can  check that, in this case, the equation
\begin{equation}
dz = - y \, dx + x \, dy
\end{equation}
does not satisfy the integrability condition \eqref{eq:Xcon3}. Nevertheless, the rays \eqref{eq:helicex} do satisfy the ray equation when
\begin{equation}
n(x,y,z) = \frac{1}{\sqrt{x^{2}+y^{2}}} \, ,
\end{equation} 
where we take the coordinates normalized in such a way that $x^{2} + y^{2} < 1$. We thus conclude that rays without orthogonal wavefronts exist in this case.

One could think that this strange behavior occurs only in very special inhomogeneous media that impart exotic rotational behaviors. This is not the case, as the following example clearly demonstrates: the vector field 
\begin{equation}
  \label{eq:ex4}
  \mathbf{X} = \begin{pmatrix}
  \displaystyle
  \frac{x}{y-1} \\
  1 \\
  \displaystyle
  \frac{z}{y}  
  \end{pmatrix} \, ,
\end{equation} 
originates the congruence
\begin{equation}
  \label{eq:msl}
  \frac{x-a}{a} = \frac{y}{-1} = \frac{z}{-b} \, ,
\end{equation}
where $a, b \in \mathbb{R}$. Here, the rays are straight lines, as we can appreciate in Figure~\ref{fig:vf}, and therefore light is propagating in a homogeneous medium, with a constant refractive index. These rays pass simultaneously by the axis $X$ and the axis $y=1$. The integrability (\ref{eq:Xcon3}) does not hold either here, so we are dealing with a rectilinear non-normal congruence.

To conclude this section, we show that for any solution $\mathbf{x} (t)$ of the ray equation, we can locally construct a vector field $\mathbf{X} (\mathbf{x})$ with local orthogonal surfaces and having $\mathbf{x} (t)$ as one of its solutions. 

In fact, let $\mathcal{C}$ be a local graph of $\mathbf{x} (t)$ ($t \in [a, b]$). $\mathcal{C}$ is a simple smooth curve in $\mathbb{R}^{3}$ (no self-intersections are allowed). Consider now the infinite family of normal planes $\pi_{\alpha}$ ($\alpha \in \mathbb{R}$) to $\mathcal{C}$ at $P \in \mathcal{C}$ and small bits $B_{\alpha}$ of them near $P=P(\alpha)$. By making these pieces sufficiently small, we can make them disjoint. Define now $\mathbf{X}$ on $B_{\alpha}$ in this way:
\begin{equation}
\mathbf{X} (P) = \left . \dot{\mathbf{x}} (t) \right |_{P} \, ,
\qquad
\mathbf{X} (B_{\alpha} \backslash P)= X_{0} (\alpha) \,,
\end{equation}
with $\mathbf{X}_{0} (\alpha)$ being any vector field extending $\left . \dot{\mathbf{x}} (t) \right |_{P} $ smoothly and orthogonal to $B_{\alpha}$ at each of its points. This local vector field $\mathbf{X}$ obviously has $B_{\alpha}$ as local orthogonal transversals\mbox{ near $\mathcal{C}$.}

\section{Rays on Level Sets of the Refractive Index}

We start by considering rays lying on the level set $n=n_{0}$, assuming that this is a surface, so that $\nabla n |_{n_{0}} \neq 0$. Writing $\ddot{\mathbf{x}} (t) = \dot{v} \bm{\tau} + v^2/\varrho \, \bm{\nu}$, with $v = \lVert \dot{\mathbf{x}} \rVert$, $\bm{\tau}$ and $\bm{\nu}$ the unitary vector tangent and normal to $\mathbf{x} (t)$ and $\varrho$ is the radius of curvature of $\mathbf{x}(t)$. Since we are in a surface, and $v^{2} = 1/n_{0}^{2}$, the ray equation gives in this case that $v$ is constant and \mbox{determined by}
\begin{equation}
\label{eq:n0con}
\frac{v^{2}}{\varrho} \bm{\nu} = \frac{\nabla n}{n^{3}} \, ,
\end{equation}
and, consequently, the graph of $\mathbf{x} (t)$ is a geodesic on $n=n_{0}$.

On the other hand, if $\lVert \nabla n \rVert$ has a constant value on $n=n_{0}$, it follows from \eqref{eq:n0con} that $\varrho = n_{0} / \lVert \nabla n \rVert$ has a constant value along $\mathbf{x} (t)$. This assumption holds when $n(\mathbf{x})$ is of one of the following forms:
\begin{enumerate}
  \item $n(x^2 + y^{2} + z^{2})$
  \item $n(x^2 + y^{2})$
  \item $n(z)$ .
\end{enumerate}

Observe that $\lVert \nabla n \rVert$ is also constant when $n$ satisfies the
eikonal equation
\begin{equation}
\lVert \nabla n \rVert^{2} = f(n)
\end{equation}
for some smooth non-negative function $f(n) > 0$.

For the three aforementioned cases, the graph of $\mathbf{x} (t)$ is as follows:
\begin{enumerate}
  \item A maximum circle on the sphere $n=n_{0}$ (which we parametrize as $x^{2} + y^{2} + z^{2} = R^{2}$) with radius
  \begin{equation}
  R = \frac{n_{0}(R^{2})}{\lvert 2 n^{\prime}_{0} (R^{2}) R \rvert} \, .
  \end{equation}

  \item A helix on the cylinder $x^{2} + y^{2} = R^{2}$ and $\varrho$ constant. The value of $\nabla n |_{x^{2} + y^{2} = R^{2}}$ implies that straight lines parallel to the $z$-axis cannot be light rays.

  \item A straight line $L$ on the plane $n(z) = n_{0}$, say $z=0$.
   \end{enumerate}

The infinity of solutions obtained in this way can be understood by noticing that the ray equation is symmetrical under rotations when $n(x^2 + y^{2} + z^{2})$, under rotations around the $z$ axis and translations along the $z$ axis when $n(x^{2} + y^{2})$ and under translations along the $x$ and $y$ axes when $n(z)$.

Consider now the case in which the level set $n= n_{0}$ is not a surface but a curve $\mathcal{C}$. In this case, it is trivial to show that $\nabla n |_{n=n_{0}} = 0$. Assume in addition that $\mathcal{C}$ is a straight line $L$ (say, the $z$ axis). Then, it is easy to check that $\mathbf{x} (t) = (0, 0, t/n_{0})^{\top}$ is a solution of the \mbox{ray equation.}

Other straight-line solutions that are global in $t$ are obtained when $\nabla n|_{L}$ is parallel to $L$. In this case, the ray equation has a solution of the form 
$\mathbf{x} (t) = (0, 0, z(t))^{\top}$, when $z(t)$ satisfies
\begin{equation}
\ddot{z} + A(z) \dot{z}^{2} = B(z) \, ,
\qquad
A(z) = \left . \frac{\partial_{z} n}{n} \right |_{z} \, , \quad
B(z) = \left . \frac{\partial_{z} n}{n^{3}} \right |_{z} \, .
\end{equation}

This equation is integrable and reduces to the linear equation
\begin{equation}
\label{eq:line}
\frac{1}{2} \frac{du}{dz} + Au = B \,,
\end{equation}
with $u = \dot{z}^{2}$. Since $n^{2} \dot{\mathbf{x}}^{2} = 1$, $0 < \dot{z}^{2}(t) = 1/n^{2} < 1 $ and so
\begin{equation}
\label{eq:jod}
\frac{dz}{dt} = \pm \sqrt{u(z,C)} \, ,
\end{equation}
where $C$ is an integration constant. We finally get
\begin{equation}
t-t_{0} = \int_{z_{0}}^{z} \frac{dz}{+ \sqrt{u(z,C)}} \ge \int_{z_{0}}^{z} dz 
\end{equation}
and therefore the solution \eqref{eq:jod} is defined for every $t\in \mathbb{R}$.

\section{Concluding Remarks}

In summary, we have challenged a traditional conviction by showing that
non-normal congruences are allowed by the laws of geometrical optics. Even if wavefront detection is becoming a crucial tool for many modern optical technologies~\cite{Geary:1995fk}, our findings seem to suggest that the role of rays, as the carriers of the observable energy, should not\mbox{ be overlooked.}

This research was funded by Ministerio de Ciencia, Innovaci\'on y Universidades Grant number PGC2018-099183-B-I00. We thank Juan J. Monz\'on. Jos\'e L. Romero and Aaron Z. Goldberg for useful discussions.

\end{document}